\label{key}
\documentclass[aip,apm,amsmath,amssymb,twocolumn,reprint]{revtex4-1}
\usepackage{graphicx}
\usepackage{bm}
\usepackage{amsmath}
\usepackage{physics}
\usepackage{mathrsfs}
\usepackage{calrsfs}
\begin{document}

\title{Magnetic circular polarization of luminescence in Bismuth-doped \mbox{silica} glass}

\author{Oleksii Laguta}
\affiliation{CERLA, PHLAM UMR CNRS 8523, University Lille 1, 59655 Villeneuve d'Ascq, France}
\author{ Hicham El Hamzaoui}
\author{Mohamed Bouazaoui}
\affiliation{IRCICA - UMR8523/FR3024 CNRS, Parc de la Haute Borne, 50 av. Halley, 59658 Villeneuve d'Ascq, France}
\author{Vladimir B. Arion}
\affiliation{Institute of Inorganic Chemistry, University of Vienna, Waehringer Str. 42, A-1090 Vienna, Austria}
\author{Igor Razdobreev}
\email{Igor.Razdobreev@univ-lille1.fr}
\affiliation{CERLA, PHLAM UMR CNRS 8523, University Lille 1, 59655 Villeneuve d'Ascq, France}
\date{\today}

\begin{abstract}
The magnetic field induced circular polarization of near infrared photoluminescence in Bi-doped pure silica glass was studied
in the spectral range of 660\,-\,1600\,nm covering three excited state levels. The highest degree of magnetic circular polarization of
luminescence was observed in the lasing, first excited state (peak emission at 1440\,nm). The results 
of variable temperature and variable magnetic field measurements allows to conclude that the near infrared luminescence originates 
from an isolated non-Kramers doublet of the even-electron system.
\end{abstract}

\maketitle

Over the past decade there has been increasing interest in the development of Bi-doped fibre lasers (BFL) and amplifiers
(BFA). Despite the significant progress achieved in recent years \cite{Bufetov09}, such devices suffer from 
a number of drawbacks. The very low levels of Bismuth doping and, as a consequence, significant fiber length (typically
80\,-\,100\,m) are necessary to ensure the efficient BFL and BFA operation. Also, the efficiency of the BFL and BFA remains
significantly lower in comparison to their rare earth counterparts. Unfortunately, the poor understanding of the nature of
 near-infrared (NIR) photoluminescence (PL) in Bi-doped glasses does not allow the development of
efficient devices. Indeed, since the first demonstration of the NIR PL in Bi-doped silica glasses \cite{Murata99,Fujimoto01}
and up to now the nature of the NIR PL in Bi-doped glasses remains a subject of debates \cite{Peng11,Dianov12}.
In this context new experiments are necessary  to clarify the nature of luminescent centers in Bi-doped silica glasses. 

It is well known \cite{Stephens1974a,Riehl1976a} that the magnetic circular dichroism (MCD) and closely related
magnetic circular polarization of luminescence (MCPL) are universal techniques for investigation of the paramagnetic
impurities when the detection of the electron spin resonance (ESR) becomes difficult or even impossible. 
Both effects are considered as a sum of the contributions from the so-called $\mathscr{A}$, $\mathscr{B}$
and $\mathscr{C}$ terms, originally introduced by R.~Serber \cite{Serber1932}. The only temperature dependent and purely
paramagnetic term $\mathscr{C}$ is due to the Boltzmann population distribution among the Zeeman-split sublevels.
The diamagnetic term $\mathscr{B}$ is due to the magnetic field (MF) mixing between a degenerate level (initial) and another one
(excited, ground or even itself). Finally, the term $\mathscr{A}$, also called diamagnetic, is caused by the difference in the 
energies of the final states (excited in MCD and \lq\lq ground\rq\rq~in MCPL).
In the present Letter we report on the MCPL in Bi-doped silica without  other 
co-dopant. The choice of the MCPL technique is due to the very low absorption in the NIR spectral region and our 
preference for the bulk samples with Bismuth content as close as possible to the core of lasing fiber. For the first time we 
show that some of the luminescence bands in the Bi-doped silica glass  are the magnetic multiplets. 
The degree of MCPL is defined as $\Delta_\text{MCPL} =(I^{+}-I^{-})/I_\text{tot}$, where
$I_\text{tot} =(I^{+}+I^{-})$, $I^{+}$ and $I^{-}$ are the intensities of $\sigma^{+}$ and $\sigma^{-}$ components, respectively.
The highest degree of MCPL was found in the luminescence from the lowest, lasing excited state (ES). At $\lambda_\text{em}$\,=\,1440\,nm 
corresponding to the NIR PL maximum at 1.48\,K  and in the MF of 6T we measured $\Delta_\text{MCPL}$\,$\approx$\,0.28.

The samples of silica glasses were produced following the technique described previously \cite{Razdobreev2010a}.
At the first step the nanoporous silica xerogel in the form of cylindrical rod with an average pore diameter of about
20\,nm was prepared using a sol-gel technique \cite{HEH10}. Then this xerogel rod was solution doped with 
the precursor [Bi(Hsal)$_3$(bipy)$\cdotp$C$_7$H$_8$]$_2$, where Hsal = O$_2$CC$_6$H$_4$-2-OH. The
synthesis of the precursor was described previously \cite{Thurston2002a}. After the dehydroxylation
procedure under chlorine/oxygen atmosphere the samples were sintered at 1300\,$^{\circ}$C under helium atmosphere
resulting in a dense, transparent and colorless monolithic cylindrical preform of 5 mm diameter. This preform was cut
and polished to produce the samples with the dimensions of  2\,$\times$\,4\,$\times$\,5\,mm$^3$. The molar ratio Bi/Si inside the 
preform was determined by the electron probe microanalysis (EPMA), and it was estimated to be 
around 300 ppm. 
\begin{figure}[htb]
\centering\includegraphics[width=8cm]{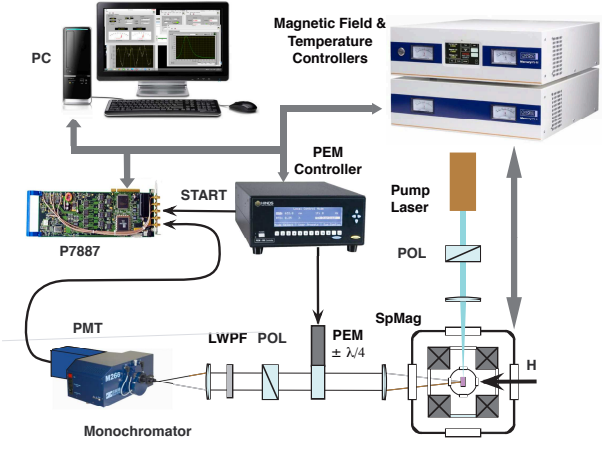}
\caption{Experimental setup. SpMag - Oxford SpectromagPT magneto-optical cryostat; 
PEM - photoelastic modulator; POL - polarizer; LWPF - long wave pass optical filter;
PMT - photomultiplier tube; P7887 - photon counting card (multiscaler).}
\label{fig:f1}
\end{figure}
The experimental setup for measurements of MCPL is schematically shown in Fig.\,\ref{fig:f1}. The experiments in the temperature 
range of 1.45\,-\,300\,K and MF's in the range of  0\,-\,7\,T were performed in the closed cycle magneto-optical
cryostat (SpectromagPT, Oxford Instr.). The thermal stability of the samples attached to the holder of the variable
temperature insert  was about of 0.01\,K except the range from 4.2 to 10\,K where  the thermal stability was
$\sim$\,0.05\,K. Laser diode or frequency doubled Ti:Sapphire laser were used for the excitation at 375\,nm.
The signal of MCPL was measured in Faraday geometry (the external MF is parallel to vector $\vec{k}$
of emitted photons). The quarter-wave retardation ($\pm\lambda/4$) was introduced by the photoelastic modulator 
(I/FS-20, Hinds Instr.) at the frequency of 20.077\,kHz. The resulting PL emission was analyzed by the fixed linear polarizer 
(LPNIR100-MP2, Thorlabs), then filtered through a monochromator and detected with the nitrogen cooled InP/InGaAs 
photomultiplier R5509-73 (Hamamatsu Inc.). The photon counting technique (P7887 scaler, Fast ComTec) was used to 
record the signal because of its better signal to noise ratio. The spectral resolution in the range 
of the first ES at 1440\,nm was $\sim$ 5\,nm and 1.5\,nm in the range of 650\,-\,950\,nm.

\begin{figure}[ht]
\centering
\includegraphics[width=8cm]{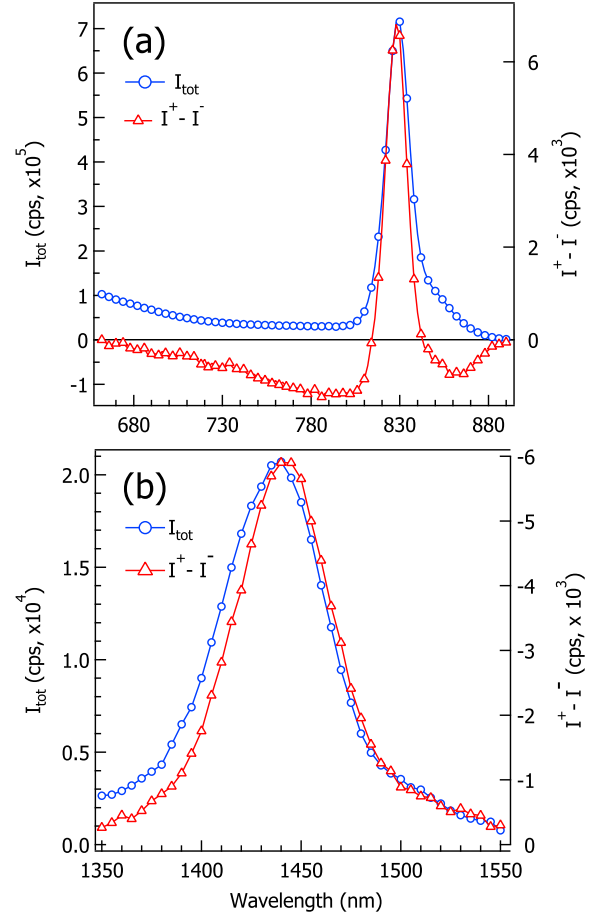}
\caption{Spectra of the total ($I_\text{tot}$) and MCPL ($I^{+}-I^{-}$)  intensities  corresponding to the third (a) and first (b) excited 
states of the luminescent center in Bi-doped silica glass.  T\,=\,1.48\,K, B\,=\,6\,T,
$\lambda_\text{exc}$=\,375\,nm, P$_\text{exc}$ = 7\,mW. (c) Magnetic field dependences of $\Delta_\text{MCPL}$ as a function of
$\mu_{B}B/\text{k}_{B}T$.  (d) Temperature dependence of $\Delta_\text{MCPL}$ at various MF's. 
Markers and solid lines correspond to experimental data and to the global fit, respectively. $\lambda_\text{det}$ = 1440\,nm.}
\label{fig:f2}
\end{figure}

The details on the photoluminescence in Bi-doped silica glass without other co-dopant at low temperature (10\,K) and at various 
excitation wavelengths were reported previously \cite{Razdobreev2010a}. Therefore we show in Fig.\,\ref{fig:f2}(a) and \ref{fig:f2}(b) 
only the comparison of luminescence and MCPL signals  ($I^{+}-I^{-}$) obtained by applying a field of 6\,T at 1.48\,K in two spectral 
ranges that correspond to the first (1350\,-\,1550\,nm) and to the third ES's (650\,-\,890\,nm). In the range of the second ES 
(900\,-\,940\,nm) we did not detect any MCPL signal. The MCPL spectrum reveals that the band around 830\,nm consists of three 
components. The most intense one exhibits positive MCPL signal,  $I^{+}-I^{-} >$\,0, with corresponding peak value of 
$\Delta_\text{MCPL}$\,=\,+0.0095 measured at 830\,nm.  Two other bands are seen in the MCPL as side bands with $I^{+}-I^{-} <$\,0. 
The MCPL band at 1440\,nm has a much simpler form with negative sign in the whole m{\tiny }easured wavelength range. The corresponding 
peak value is  $\Delta_\text{MCPL}=-$0.28 in the MF of 6\,T at 1.48\,K. Both MCPL bands (the principal in the case 
of 830\,nm band), are  blue- and red-shifted compared to  the corresponding luminescence bands (see Fig.\,\ref{fig:f2}). Moreover, this 
shift  is more pronounced on the one side, short-wavelength for the band at 1440\,nm and long-wavelength for the band at 830\,nm.
It is straightforward to show that at very low temperatures and high fields such behavior can be assigned to the contribution in
$\Delta_\text{MCPL}$ of the term which has the linear spectral dependence. The only term exhibiting a spectral dependence is the
$\mathscr{A}$-term which for the case of  MCPL is due  to the transitions to Zeeman components of the GS \cite{Zapasskii1975}. 
The $\mathscr{A}$-term  contribution should be approximately zero at the spectral band maximum \cite{Zapasskii1975,Schatz1978a}
that allows to simplify significantly the investigation of the effects of MF and temperature. Below we consider the field and temperature 
dependences of $\Delta_\text{MCPL}$ only at the peak wavelength of the total luminescence $I_\text{tot}$ of the first ES at 1440\,nm.
\begin{figure*}[ht]
\centering
\includegraphics[width=15cm]{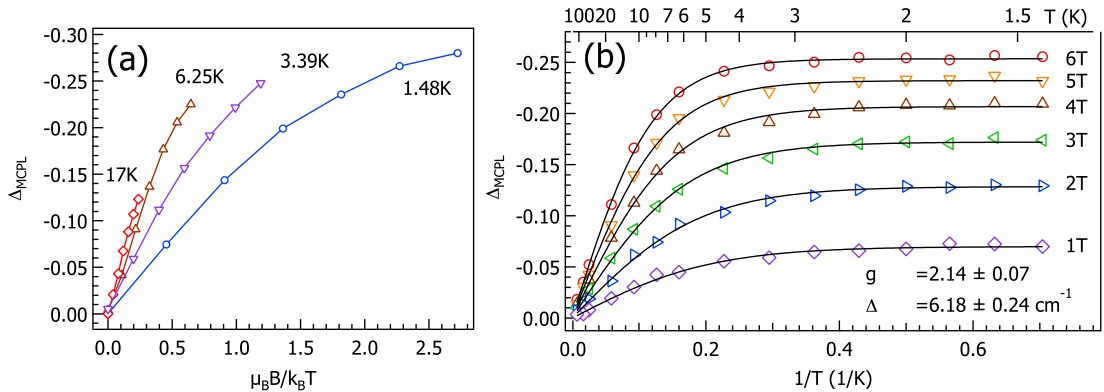}
\caption{a) Magnetic field dependences of $\Delta_\text{MCPL}$ as a function of
$\mu_{B}B/\text{k}_{B}T$.  d) Temperature dependence of $\Delta_\text{MCPL}$ at various MF's. 
Markers and solid lines correspond to experimental data and to the global fit, respectively. $\lambda_\text{det}$ = 1440\,nm.}
\label{fig:f3}
\end{figure*}
Fig.\,\ref{fig:f3}(a) shows the MF dependences of $\Delta_\text{MCPL}$ plotted as a function of $\mu_{B}B/k_{B}T$ at 
fixed temperatures, where $\mu_{B}$ is the Bohr magneton, and $k_{B}$ is the Boltzmann constant. It is seen that there is no 
superposition and the isotherms exhibit nesting behavior. Fig.\,\ref{fig:f3}(b) displays the temperature dependences of $\Delta_\text{MCPL}$
plotted as a function of $1/T$ at fixed MF's. It is seen that the saturation values of $\Delta_\text{MCPL}$ are different for each 
field. Magnetic saturation occurs within a paramagnetic state when the higher lying levels are depopulated due to the field splitting or due to
the temperature decrease. For an isolated Kramers doublet (systems with unpaired electron)  we expect to observe the field 
independent limit for the $\Delta_\text{MCPL}$ and the unique curve when the latter is plotted as a function of $\mu_{B}B/k_{B}T$.
If there is an additional level (from another Kramers doublet, for example) in a close  proximity to the lowest one, then due to 
the field-induced mixing the $\mathscr{B}$-term appears and can result in the behavior similar to that shown in Fig.\,\ref{fig:f2}(a) but 
the character of the saturation curves plotted as a function of $1/T$ remains very different from that shown in Fig.\,\ref{fig:f2}(b). The same 
conclusion follows from a consideration of the case when the transition contains the so-called $z$-polarization \cite{Solomon1995a} 
and the effects of both $g_\perp$ and $M_z/M_{xy}$, the ratio of the $z$-polarized to the $xy$-polarized transition dipole moments, 
are taken into account.

\begin{figure}[htb]
\centering\includegraphics[width=8cm]{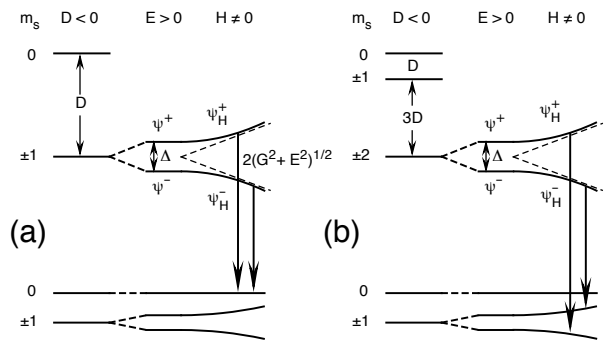}
\caption{Energy level diagram to explain the temperature and field dependent effects in MCPL. (a) Spin-triplet ground state and isolated non-Kramers doublet 
from the spin-triplet S\,=\,1 excited state; (b) Same as in (a), but  ES is spin-quintet. All splittings are exaggerated. See text. }
\label{fig:f4}
\end{figure}

On the other hand, for the systems with a zero-field splitting (non-Kramers, even electron systems), the changes in 
the population of higher levels should affect the MCPL intensity  leading to the nesting of saturation magnetization
curves taken at different temperatures in accordance with Fig.\,\ref{fig:f3}(a). Alternatively, plotting the temperature dependence at constant field 
separates the temperature and field effects as it is shown in Fig.\,\ref{fig:f3}(d). Fig.\,\ref{fig:f4} illustrates two non-exhaustive examples of the 
energy level diagrams corresponding to this kind of systems.  Both examples contain all necessary ingredients to explain all the observed features 
in the behavior of $\Delta_\text{MCPL}$. The ES of the system shown in Fig.\,\ref{fig:f4}(a) is a spin-triplet state and the GS is also a spin triplet.  
Though we consider below only this particular example, the result is identical to the case of the spin-quintet ES show in Fig.\,\ref{fig:f4}(b). 
The analysis of  the saturation curves for such a system can be performed in terms of the spin Hamiltonian with the axial (D) and rhombic (E) 
zero-field splitting parameters \cite{Abragam}:
\begin{align}
\mathscr{H}=D\left\lbrace S_{z}^2 - \frac{1}{3}S\left( S+1\right) \right\rbrace + E\left( S_{x}^2-S_{y}^2\right) 
+g_{\parallel}\mu_{B}BS_{z}\cos\theta, 
\end{align}
\noindent where $\theta$ is the angle between the MF and the axis of symmetry ($z$-axis),  and which implies $g_\perp\!\ll~g_{\parallel}$.

First, the triplet ES is split by the axial component D\,$<$\,0 that brings the doublet m$_{s}$\,=\,$\pm$1 in lowest position.
The rhombic component further removes the degeneracy ($\Delta$\,=\,2E), and the condition $\Delta\ll\abs{\text{D}}$ ensures its 
isolated character. Without an external MF the spin Hamiltonian is diagonalized with the real wavefunctions
$\psi^{\pm}=1/\sqrt{2}\qty(\ket{1}\pm\ket{\bar{1}}) $.  It is clear that the angular momentum is quenched and there is no circularly polarized
emission. The MF appears as an off-diagonal pure imaginary perturbation $g\mu_{B}B\cos\theta$. This Zeeman perturbation changes the 
wavefunctions  and now they take the form: $\psi^{+}_H=\cos\alpha\ket{1} + \sin\alpha\ket{\bar{1}}$ and $\psi^{-}_H=\sin\alpha\ket{1} - \cos\alpha\ket{\bar{1}}$,  
where $\tan2\alpha$\,=\,$\Delta$/2G and G\,=\,$g_{\parallel}\mu_{B}B\cos\theta$ \cite{Abragam}.

The described behavior 
is quite similar to that observed in MCD for the isolated non-Kramers doublet \cite{Solomon1995a}. To use the corresponding 
analytical expression derived for the case of MCD it must be corrected for MCPL by taking into account the effect of photoselection 
according to Schatz et al. \cite{Schatz1978a}. The resulting formula  which describes the temperature and MF dependences 
of $\Delta_\text{MCPL}$ in a disordered system is:
\begin{equation}
\Delta_\text{MCPL} \\ 
=A_{sat}\int_0^1\!\frac{2n^{4}G_z}{\sqrt{\Delta^{2}+4n^{2}G_z^{2}}}\tanh\left( \frac{\sqrt{\Delta^{2}+4n^{2}G_z^{2}}}{2kT}\right)\mathrm{d}n,  
\end{equation}
\noindent where $G_z  = g_{\parallel}\mu_{B}B$ and $n =\cos\theta$. This equation is valid to describe the behavior of the MCPL
from the  strongly  anisotropic non-Kramers doublet with $g_\perp\!\ll\!g_{\parallel}$ and $M_z/M_{xy}=0$. The nonzero $g_\perp$ 
and $M_z/M_{xy}$ could be taken into account  also, however, the additional parameters will soften the overall fit without affecting 
the main conclusions. The simultaneous fit of all curves shown in Fig.\,\ref{fig:f2}(d) by simulated annealing method \cite{Press1992}
results in $g_\parallel$=2.14$\pm$0.07 and $\Delta$\,=\,6.18$\pm$0.24\,cm$^{-1}$. The obtained value of g-factor indicates that the
first excited can not be put in correspondence to the $^3$P$_1$ state of the isolated Bi$^+$ ion for which we expect $g\leq1.5$.  
It must be emphasized that the proposed energy level diagrams are non-exhaustive because our experiment does not reveal neither 
the exact multiplicity nor the character of the lowest spin component of the GS. Nevertheless, by simple evaluation of the population 
ratio in the ES we can conclude that the contribution of the $\mathscr{C}$-term in the described above spectral shift of MCPL band 
is negligible. Then this shift is due to the $\mathscr{A}$-term, as a consequence, the GS of the luminescent center is a magnetic 
multiplet and this is the second important finding in our experiment.  This again eliminates isolated Bi$^+$ ion as a possible 
luminescent center. It worth noting here, that regardless what spin sublevel of the GS is lowest, in the frame of the presented 
model the ESR is inaccessible. For instance, if  $m_s$\,=\,$\pm$1 (or $m_s$\,=\, $\pm$2) is lowest sublevel of GS, one can expect 
that  the condition $\abs{\text{D}}\gg\Delta$ is also valid while the splitting of the GS doublet  is at least comparable to the ES one. 
This explains the fact that even \lq\lq forbidden\rq\rq~electron spin resonance with $\Delta m_s$\,=\,$\pm$2
in Bi-doped glasses could not  be detected in X-band (the photon energy $\approx$0.33 cm$^{-1}$).

Sokolov et al.~\cite{Sokolov2008,Sokolov2015} proposed first Bi$_{2}^{-}$ and Bi$_{2}^{2-}$ dimers then the complexes
of divalent bismuth with neutral oxygen vacancies \mbox{Bi$^{2+}$\,-V$_\text{O}$} to explain the NIR PL observed in Bi-doped 
glasses. It should  be clear from the above discussion that Bi$_{2}^{-}$ dimers with unpaired electron cannot be responsible 
for the observed  temperature dependence of MCPL. In the recently proposed model of \mbox{Bi$^{2+}$\,-V$_\text{O}$} the first 
ES is a  perfectly isolated Kramers doublet  (see Fig.\,3 in Ref.~\cite{Sokolov2015}). Thus, this model is also 
inconsistent with  our experiment. In the model of Bi$_{2}^{2-}$ all emitting states are singlet while the GS is a triplet 
(see Fig.\,2 in Ref.~\cite{Sokolov2008}). In such a system only $\mathscr{A}$-term can appear in the MCPL. However, 
this term is temperature independent and it can exhibit only the linear field dependence~\cite{Zapasskii1975}. It follows that the 
model  of isolated Bi$_{2}^{2-}$ dimers is also inconsistent with our experimental observations. In all above mentioned models 
the GS is a spin multiplet that implies that  ESR in principle can be observed. Though Khonthon et al.~\cite{Khonthon2007} 
reported on the observation of an ESR signal with $g$\,$\approx$2.2, its assignment to some  luminescent center cannot  
be unambiguous without additional experiments. On the contrary, our experiment  directly connects the particular
luminescence band to the paramagnetic properties of the luminescent center. Finally,  Dianov~\cite{Dianov2010} among other 
possible origins of NIR PL in Bi-doped silica suggested Bi$^{2+}$-Bi$^{3+}$ dimer and Bi$^{+}$ ion between two neutral vacancies. 
If the first one is again the system with unpaired electron, the latter one should be considered at least in the frame of molecular 
orbitals because the GS of the isolated Bi$^+$ ion is a singlet.

In conclusion, MCPL in Bi-doped silica glass was investigated in a wide temperature and MF ranges. 
It was found that the photoluminescence from the first ES (lasing level) exhibits a strong magnetic circular
polarization. The experiments put in evidence the spin multiplicity of the ES and GS of the luminescent 
center. The field and temperature dependences of the MCPL signal recorded at 1440\,nm rule out the 
odd-electron system as a possible origin of the NIR photoluminescence.  This correlates well with the known 
experimental fact that ESR cannot be observed neither in the GS nor in the ES with the standard 
method. The results of the experiment can be explained in the assumption that the NIR PL originates from an isolated 
non-Kramers doublet (ES) and the GS is the spin-triplet or spin-quintet state of the even-electron luminescent center.

\section*{Funding Information}
 \lq\lq Agence Nationale de la Recherche\rq\rq,~grant  ANR  \mbox{\lq\lq BOATS\rq\rq}~12BS04-0019-01.

\section*{Acknowledgments}

I.R. is grateful to L.\,F.\,Chibotaru for reading the manuscript and his comments.

\end{document}